\documentclass[aps,preprint,showpacs,showkeys,nofootinbib,floatfix]{revtex4}
\usepackage{epsfig}
\usepackage{graphicx}
\begin{document} 

\title{\textbf{Nonexistence of a $\Lambda nn$ bound state}}
\author{H.~Garcilazo} 
\email{humberto@esfm.ipn.mx} 
\affiliation{Escuela Superior de F\' \i sica y Matem\'aticas, \\ 
Instituto Polit\'ecnico Nacional, Edificio 9, 
07738 M\'exico D.F., Mexico} 

\author{A.~Valcarce} 
\email{valcarce@usal.es} 
\affiliation{Departamento de F\'\i sica Fundamental,\\ 
Universidad de Salamanca, E-37008 Salamanca, Spain}

\date{\today} 

\begin{abstract} 
It has been recently suggested the existence of a neutral bound state
of two neutrons and a $\Lambda$ hyperon, $^3_\Lambda n$.
We point out that using either simple separable potentials or
a full-fledged calculation with realistic 
baryon-baryon interactions derived from the constituent quark cluster
model there is no possibility for the existence of such a $\Lambda nn$
bound state. For this purpose, we performed a full Faddeev calculation of the $\Lambda nn$ system 
in the $(I,J^P)=(1,1/2^+)$ channel using the interactions
derived from the constituent quark cluster model which describe well the
two-body $NN$ and $NY$ data and the $\Lambda np$ hypertriton.
\end{abstract}

\maketitle 

In a recent Rapid Communication by the experimental HypHI Collaboration~\cite{HYPHI} 
it has been suggested the existence of a neutral bound state
of two neutrons and a $\Lambda$ hyperon, $^3_\Lambda n$. They analyze the
experimental data obtained from the reaction $^6$Li +$^{12}$C at 2A GeV
to study the invariant mass distribution of $d+\pi^-$ and $t+\pi^-$.
The signal observed in the invariant mass distributions of $d+\pi^-$ and $t+\pi^-$ 
final states was attributed to a strangeness-changing weak process corresponding
to the two- and three-body decays of an unknown bound state of two neutrons associated
with a $\Lambda$, $^3_\Lambda n$, via $^3_\Lambda n \to t + \pi^-$ and
$^3_\Lambda n \to t^* + \pi^- \to d + n + \pi^-$.

This is an intriguing conclusion since one would naively expect the $\Lambda nn$ system
to be unbound. In the $\Lambda nn$ system the two neutrons interact in the $^1S_0$
partial wave while in the $\Lambda np$ system they interact in the $^3S_1$ partial wave.
Thus, since the $NN$ interaction in the $^1S_0$ channel is weaker than the $^3S_1$ channel,
and the $\Lambda np$ system is bound by only 0.13 MeV, one may have anticipated that the
$\Lambda nn$ system should be unbound. The unbinding of the $\Lambda nn$ system was first 
demonstrated by Dalitz and Downs~\cite{Dal58} using a variational approach. 

In a previous, by now somewhat older, paper~\cite{GAJPG} we concluded the 
nonexistence of $\Lambda nn$ bound states by solving 
the Faddeev equations with separable potentials whose parameters were 
adjusted to reproduce the $\Lambda n$ scattering length and effective 
range of the two-body channels as obtained from four different 
versions of the Niemegen model~\cite{NIEM1,NIEM2,NIEM3,NIEM4} as well
as the corresponding $NN$ spin-singlet and spin-triplet low-energy parameters. This 
leads to integral equations in one continuous variable.

As pointed out in Ref. \cite{GAJPG}, if a system can have at most 
one bound state then the simplest way to determine if it is bound or 
not is by looking at the Fredholm determinant $D_F(E)$ at zero 
energy. If there are no interactions then $D_F(0)=1$, 
if the system is attractive then $D_F(0)<1$, and if a
bound state exists then $D_F(0)<0$. We found in Ref.~\cite{GAJPG}
that $D_F(0)$ lies between 0.46 and 0.59 for the different models
constructed by the Niemegen group so that the system is quite 
far from being bound.

Of course, it can be argued that the use of simple separable potentials 
is not a realistic assumption. Besides, since our previous work the knowledge
of the strangeness --1 two-baryon system has improved and the models to 
study these systems are more tightly constrained. Therefore, 
we have now reexamined the $\Lambda nn$ system within a 
realistic baryon-baryon formalism obtained from the quark
model. The baryon-baryon interactions involved in the study of the coupled
$\Sigma NN - \Lambda NN$ system are obtained from the 
constituent quark cluster model~\cite{Val05,Gar05}. In this model baryons
are described as clusters of three interacting massive (constituent) quarks,
the mass coming from the spontaneous breaking of chiral symmetry. The
first ingredient of the quark-quark interaction is a confining
potential. Perturbative aspects of QCD are taken into account
by means of a one-gluon potential. Spontaneous breaking of 
chiral symmetry gives rise to boson exchanges
between quarks. In particular, there appear pseudoscalar 
boson exchanges and their corresponding scalar partners~\cite{GFV07,GAXXX}.
Explicit expressions of all the
interacting potentials and a more detailed discussion 
of the model can be found in Refs.~\cite{Gar05,GFV07}.

In Refs.~\cite{GFV07,GAXXX} we established the formalism
to study the $\Lambda NN$ system at threshold using the
baryon-baryon interactions obtained from the constituent quark cluster
model which leads to
integral equations in the two continuous variables $p$ and $q$, where $p$
is the relative momentum of the pair and $q$ is the relative momentum of
the third particle with respect to the pair.
In order to solve these equations
the two-body $t-$matrices are expanded in terms of Legendre 
polynomials leading to integral equations in only one continuous 
variable coupling the various Legendre components required for
convergence.

\begin{table}[b]
\caption{Two-body $\Sigma N$ channels with a nucleon as spectator 
$(\ell_\Sigma s_\Sigma j_\Sigma i_\Sigma\lambda_\Sigma J_\Sigma)_N$,
two-body $\Lambda N$ channels with a nucleon as spectator 
$(\ell_\Lambda,s_\Lambda j_\Lambda i_\Lambda\lambda_\Lambda J_\Lambda)_N$,
two-body $NN$ channels with a $\Sigma$ as spectator 
$(\ell_N,s_N j_N i_N \lambda_N J_N)_\Sigma$, and
two-body $NN$ channels with a $\Lambda$ as spectator 
$(\ell_N s_N j_N i_N \lambda_N J_N)_\Lambda$ that contribute to
the $(I,J^P)=(1,1/2^+)$ state. $\ell$, $s$, $j$, and $i$, are, respectively, 
the orbital angular momentum, spin, total angular momentum, and isospin
of a pair, while $\lambda$ 
and $J$  are the orbital angular momentum of the third particle
with respect to the pair and the result of coupling $\lambda$ with the
spin of the third particle.}
\begin{ruledtabular} 
\begin{tabular}{cccc} 
 
$(\ell_\Sigma s_\Sigma j_\Sigma i_\Sigma\lambda_\Sigma J_\Sigma)_N$ &
$(\ell_\Lambda,s_\Lambda j_\Lambda i_\Lambda\lambda_\Lambda J_\Lambda)_N$
& $(\ell_N,s_N j_N i_N \lambda_N J_N)_\Sigma$
& $(\ell_N s_N j_N i_N \lambda_N J_N)_\Lambda$ \\
\hline
 
 (000$\frac{1}{2}$0$\frac{1}{2}$),(011$\frac{1}{2}$0$\frac{1}{2}$),  
& (000$\frac{1}{2}$0$\frac{1}{2}$),(011$\frac{1}{2}$0$\frac{1}{2}$),  
& (00010$\frac{1}{2}$),(01100$\frac{1}{2}$), & (00010$\frac{1}{2}$), \\ 
 (211$\frac{1}{2}$0$\frac{1}{2}$),(011$\frac{1}{2}$2$\frac{3}{2}$),  
& (211$\frac{1}{2}$0$\frac{1}{2}$),(011$\frac{1}{2}$2$\frac{3}{2}$),  
&  (21100$\frac{1}{2}$),(01102$\frac{3}{2}$), & \\ 
 (211$\frac{1}{2}$2$\frac{3}{2}$),(000$\frac{3}{2}$0$\frac{1}{2}$),  
& (211$\frac{1}{2}$2$\frac{3}{2}$)
& (21102$\frac{3}{2}$) &  \\ 
 (011$\frac{3}{2}$0$\frac{1}{2}$),(211$\frac{3}{2}$0$\frac{1}{2}$),  
& & & \\
 (011$\frac{3}{2}$2$\frac{3}{2}$),(211$\frac{3}{2}$2$\frac{3}{2}$)  
& & & \\
\end{tabular}
\end{ruledtabular} 
\label{t1}
\end{table}

This model takes into account the coupling $N\Lambda-N\Sigma$
as well as the tensor force responsible for the coupling between
S and D waves. In particular, for the $\Lambda NN$ channel 
$(I,J^P)=(1,1/2^+)$ which corresponds to the conjectured $\Lambda nn$
bound state there is a total of 21 coupled channels contributing
to the state. We give in Table~\ref{t1} the quantum numbers of these contributing
channels.

In Ref.~\cite{GAXXX} we showed that if one increases the triplet
$N\Lambda$ interaction by increasing the triplet scattering length
then the $\Lambda NN$ state with $(I,J^P)=(0,3/2^+)$ becomes
bound and since that state does not exist we are allowed to 
set an upper limit of 1.58 fm for the $\Lambda N$ spin triplet
scattering length. Since, in addition, the fit of the 
hyperon-nucleon cross sections is worsened~\cite{GFV07}
when the spin-triplet
scattering length is smaller than 1.41 fm we concluded that
$1.41 \le  a_{1/2,1} \le 1.58$ fm. By requiring that the 
hypertriton binding energy had the experimental value $B=0.13\pm 0.05$
MeV we obtained for the $\Lambda N$ spin-singlet scattering length
the limits $2.37 \le a_{1/2,0} \le 2.48$ fm. 

Thus, we constructed twelve
different models corresponding to different choices of the spin-singlet 
and spin-triplet $\Lambda N$ scattering lengths which describe equally well all the 
available experimental data. We solved the three-body problem taking full account
of the $\Lambda NN - \Sigma NN$ coupling as well as the effect of the D waves.
We present in Table~\ref{t2} the Fredholm determinant at zero energy of the $(I,J^P)=(1,1/2^+)$
state for these models. The realistic quark model interactions predict
a Fredholm determinant at zero energy ranging between 0.38 and 0.42, close to the
interval 0.46--0.59 obtained from the separable potentials of the Niemegen group.
As one can see, in all cases the Fredholm determinant
at zero energy is positive and far from zero, excluding the possibility
for binding in this system. From the results of
Table~\ref{t2} and from the energy dependence of the Fredholm
determinant shown in Fig. 2 of Ref.~\cite{GAXXX} one can infer that
the $(I,J^P)=(1,1/2^+)$ state is unbound by at least 5--10 MeV, which is a large
energy in comparison with the 0.13 MeV binding energy of the hypertriton.

\begin{table}[b]
\caption{Fredholm determinant at zero energy $D_F(0)$ for
several hyperon-nucleon interactions characterized by
$\Lambda N$ scattering lengths $a_{1/2,0}$ and $a_{1/2,1}$
(in fm).}
\begin{ruledtabular} 
\begin{tabular}{ccccc} 
  & $a_{1/2,1}=1.41$ & $a_{1/2,1}=1.46$ & $a_{1/2,1}=1.52$ 
& $a_{1/2,1}=1.58$   \\
\hline
 $a_{1/2,0}=2.33$ & 0.42 &  0.41 &  0.40 &  0.38 \\
 $a_{1/2,0}=2.39$ & 0.42 &  0.41 &  0.39 &  0.38 \\
 $a_{1/2,0}=2.48$ & 0.42 &  0.41 &  0.40 &  0.38 \\
\end{tabular}
\end{ruledtabular}
\label{t2} 
\end{table}

To summarize, we have shown that using either simple separable potentials or
a full-fledged calculation with realistic 
baryon-baryon interactions derived from the constituent quark cluster
model there is no possibility for the existence of a $\Lambda nn$
bound state. Thus, the signal observed in the invariant mass distributions 
of $d+\pi^-$ and $t+\pi^-$ final states in the analysis of the
experimental data obtained from the reaction $^6$Li +$^{12}$C at 2A GeV
and adduced to the existence of a neutral bound state
of two neutrons and a $\Lambda$ hyperon must be due to 
a different effect.

\acknowledgments 
This work has been partially funded by COFAA-IPN (M\'exico), 
by Ministerio de Educaci\'on y Ciencia and EU FEDER under 
Contract No. FPA2010-21750-C02-02 and by the
Spanish Consolider-Ingenio 2010 Program CPAN (CSD2007-00042).

\end{document}